\documentclass[12pt]{article}%
\usepackage{amsmath,latexsym}
\usepackage{graphicx}
\usepackage{amsmath}
\usepackage{amsfonts}
\usepackage{amssymb}%
\begin{document}

\title{{Macroscopic noncommutative-geometry wormholes
   as emergent phenomena in $f(Q)$ gravity}}
   \author{
Peter K. F. Kuhfittig*\\  \footnote{kuhfitti@msoe.edu}
 \small Department of Mathematics, Milwaukee School of
Engineering,\\
\small Milwaukee, Wisconsin 53202-3109, USA}

\date{}
 \maketitle

\begin{abstract}\noindent
Noncommutative geometry, an offshoot of string
theory, replaces point-like particles by smeared
objects.  These local effects have led to
wormhole solutions in a semiclassical setting,
but it has also been claimed that the noncommutative
effects can be implemented by modifying only the
energy momentum tensor in the Einstein field
equations, while leaving the Einstein tensor
unchanged.  The implication is that
noncommutative-geometry wormholes could be
macroscopic.  This result can be readily 
explained by considering the 
noncommutative-geometry background to be
a fundamental property and the macroscopic
wormhole spacetime to be emergent, according 
to an earlier version of this paper.  However,
it is shown in the present version that such 
wormholes could not be sufficiently massive 
to exist on a macroscopic scale.  A major
objective in this paper is to invoke $f(Q)$ 
gravity to provide the extra degrees of 
freedom to overcome these obstacles. 

\end{abstract}

\section{Introduction}\label{E:introduction}
Wormholes are handles or tunnels in spacetime
connecting widely separated regions of our
Universe or entirely different universes.  While
wormholes are as good a prediction of Einstein's
theory as black holes, they are subject to some
severe restrictions.  In particular, holding a
wormhole open requires a violation of the null
energy condition from quantum field theory,
calling for the existence of ``exotic matter"
\cite{MT88}.  This violation is more of a
practical than conceptual problem, as
illustrated by the Casimir effect \cite{hC48}:
exotic matter can be made in the laboratory.
Being a rather small effect, it may not be
sufficient for supporting a macroscopic wormhole.
That is the main issue discussed in this paper.

Before continuing, let us recall that the need
for exotic matter has always been viewed as a
serious barrier for constructing a wormhole.
Many attempts have therefore been made to reduce
the amount \cite{pK02} or to eliminate exotic
matter entirely by appealing to various modified
theories of gravity, such as $f(R)$ or $f(R, T)$
modified gravity, or even by assuming the
existence of higher dimensions.  For a detailed
discussion of this problem, see Ref. \cite{JBG},
which deals with wormholes in 4$D$
Einstein-Gauss-Bonnet gravity.

Another area dealing with the small effects
mentioned above is noncommutative geometry, an
offshoot of string theory, where point-like
particles are replaced by smeared objects, to
be discussed further below.  It has been
suggested that a noncommutative-geometry
background does not prevent a wormhole from
being macroscopic.  The reason for this outcome is
further elaborated on in Section
\ref{S:Conclusion}: the noncommutative-geometry
background is considered to be a fundamental
theory, while the resulting macroscopic scale
is an emergent phenomenon.

As noted in the Abstract, noncommutative-geometry
wormholes could not be sufficiently massive to 
exist on a macroscopic scale.  We will see in 
Sec. \ref{S:f(Q)} that by invoking $f(Q)$ gravity,
we can overcome these obstacles.

\emph{Remark:} Noncommutative-geometry wormholes
based on the Casimir effect are discussed in Ref.
\cite{pK23}.

\section{Background}\label{S:background}

\subsection{Morris-Thorne wormholes}
With the Schwarzschild solution in mind,
Morris and Thorne \cite{MT88} proposed the
following static and spherically symmetric
line element for a wormhole spacetime:
\begin{equation}\label{E:line1}
ds^{2}=-e^{2\Phi(r)}dt^{2}+\frac{dr^2}{1-b(r)/r}
+r^{2}(d\theta^{2}+\text{sin}^{2}\theta\,
d\phi^{2}),
\end{equation}
using units in which $c=G=1$.  In the now
customary terminology, $\Phi=\Phi(r)$ is called
the \emph{redshift function}, which must be finite
everywhere to prevent the occurrence of an event
horizon. The function $b=b(r)$ is called the
\emph{shape function} since it determines the
spatial shape of the wormhole when viewed, for
example, in an embedding diagram.  For our
purposes, the most important property is
$b(r_0)=r_0$, where $r=r_0$ is the radius
of the \emph{throat} of the wormhole.  According
to Ref. \cite{MT88}, the definition of throat
requires the \emph{flare-out condition}
$b'(r_0)<1$, while $b(r)<r$ for $r>r_0$ near
the throat.  The flare-out condition can only
be met by violating the null energy condition
(NEC)
\begin{equation}
   T_{\alpha\beta}k^{\alpha}k^{\beta}\ge 0
\end{equation}
for all null vectors $k^{\alpha}$, where
$T_{\alpha\beta}$ is the energy momentum tensor.
Matter that violates the NEC is called ``exotic"
in Ref. \cite{MT88}.  Applied to a wormholes
setting, observe that for the radial outgoing
null vector $(1,1,0,0)$, the violation reads
$T_{\alpha\beta}k^{\alpha}k^{\beta}=\rho+
p_r<0$.  Here $T^t_{\phantom{tt}t}=-\rho(r)$
is the energy density, $T^r_{\phantom{rr}r}=
p_r(r)$ is the radial pressure, and
$T^\theta_{\phantom{\theta\theta}\theta}=
T^\phi_{\phantom{\phi\phi}\phi}=p_t(r)$
is the lateral (transverse) pressure.  Our
final requirement is \emph{asymptotic flatness:}
$\text{lim}_{r\rightarrow\infty}\Phi(r)=0$ and
$\text{lim}_{r\rightarrow\infty}b(r)/r=0$.

For later reference, we now state  the Einstein
field equations:
\begin{equation}\label{E:E1}
  \rho(r)=\frac{b'}{8\pi r^2},
\end{equation}
\begin{equation}\label{E:E2}
   p_r(r)=\frac{1}{8\pi}\left[-\frac{b}{r^3}+
   2\left(1-\frac{b}{r}\right)\frac{\Phi'}{r}
   \right],
\end{equation}
and
\begin{equation}\label{E:E3}
   p_t(r)=\frac{1}{8\pi}\left(1-\frac{b}{r}
   \right)
   \left[\Phi''-\frac{b'r-b}{2r(r-b)}\Phi'
   +(\Phi')^2+\frac{\Phi'}{r}-
   \frac{b'r-b}{2r^2(r-b)}\right].
\end{equation}

\subsection{Noncommutative geometry}
\label{SS:NG}
The noncommutative-geometry background
mentioned in the Introduction is based
on the realization that coordinates may
become noncommutative operators on a
$D$-brane \cite{eW96, SW99}.  A critical
feature is that noncommutativity replaces
point-like particles by smeared objects
\cite{SS03, NSS06, NS10}, thereby eliminating
the divergences that are normally unavoidable
in general relativity.  According to Ref.
\cite{NSS06}, this objective can be realized
by showing that spacetime can be encoded in the
commutator $[\textbf{x}^{\mu},\textbf{x}^{\nu}]
=i\theta^{\mu\nu}$, where $\theta^{\mu\nu}$ is
an antisymmetric matrix that determines the
fundamental cell discretization of spacetime
in the same way that Planck's constant $\hbar$
discretizes phase space.  According to Refs.
\cite{NM08, LL12}, the smearing can be modeled
using a so-called Lorentzian distribution of
minimal length $\sqrt{\gamma}$ instead of the
Dirac delta function: the energy density of
a static and spherically symmetric and
particle-like gravitational source is
given by
\begin{equation}\label{E:rho}
  \rho (r)=\frac{m\sqrt{\gamma}}
     {\pi^2(r^2+\gamma)^2}.
\end{equation}
The usual interpretation is that the
gravitational source causes the mass $m$ of
a particle to be diffused throughout the
region of linear dimension $\sqrt{\gamma}$
due to the uncertainty.

Another critical aspect of noncommutative
geometry in a wormhole setting is discussed
in Ref. \cite{NSS06}: it is possible to
implement the noncommutative effects in the
Einstein field equations $G_{\mu\nu}
=\frac{8\pi G}{c^4}T_{\mu\nu}$ by modifying
only the energy momentum tensor, while
leaving the Einstein tensor $G_{\mu\nu}$ intact.
According to Ref. \cite{NSS06}, the basic reason
for this is that a metric field is a geometric
structure defined over an underlying manifold
whose strength is determined by its curvature,
but this is nothing more than the response to
the presence of a mass-energy distribution.
What is critical here, according to Ref.
\cite{NSS06}, is that noncommutativity is
an intrinsic property of spacetime rather
than some kind of superimposed structure.
So it has a direct effect on the mass-energy
and momentum distributions.  The concomitant
determination of the spacetime curvature then
explains why the Einstein tensor can be left
unchanged.  As a consequence, the length
scales could be macroscopic, to be confirmed 
in the next section.

\section{The macroscopic throat size}
Our first task in this section is to determine
the shape function based on Eqs. (\ref{E:E1})
and (\ref{E:rho}):
\begin{multline}\label{E:shape}
   b(r)=r_0+\int^r_{r_0}8\pi(r')^2\rho(r')dr'\\
   =\frac{4m}{\pi}
  \left[\text{tan}^{-1}\frac{r}{\sqrt{\gamma}}
  -\sqrt{\gamma}\frac{r}{r^2+\gamma}-
  \text{tan}^{-1}\frac{r_0}{\sqrt{\gamma}}
  +\sqrt{\gamma}\frac{r_0}{r_0^2
  +\gamma}\right]+r_0\\
  =\frac{4m}{\pi}\frac{1}{r}
  \left[r\,\text{tan}^{-1}\frac{r}{\sqrt{\gamma}}
  -\sqrt{\gamma}\frac{r^2}{r^2+\gamma}-
  r\,\text{tan}^{-1}\frac{r_0}{\sqrt{\gamma}}
  +\sqrt{\gamma}\frac{r_0r}{r_0^2
  +\gamma}\right]+r_0.
\end{multline}
Observe that $\text{lim}_{r\rightarrow
\infty}b(r)/r=0$; to ensure asymptotic
flatness, we retain the assumption
$\text{lim}_{r\rightarrow\infty}\Phi(r)=0$.
It turns out to be advantageous to let $B=
b/\sqrt{\gamma}$ be the form of the shape function
even though $B(r_0)\neq r_0$.  The reason is that
$B$ can be expressed as a function of
$r/\sqrt{\gamma}$ by a simple algebraic
rearrangement:
\begin{multline}\label{E:shape}
   \frac{1}{\sqrt{\gamma}}
   \,b(r)=
   B\left(\frac{r}{\sqrt{\gamma}}\right)=\\
   \frac{4m}{\pi}\frac{1}{r}\left[\frac{r}{\sqrt{\gamma}}
   \,\text{tan}^{-1}\frac{r}{\sqrt{\gamma}}
   -\frac{\left(\frac{r}{\sqrt{\gamma}}\right)^2}
   {\left(\frac{r}{\sqrt{\gamma}}\right)^2+1}
  -\frac{r}{\sqrt{\gamma}}\,
  \text{tan}^{-1}\frac{r_0}{\sqrt{\gamma}}
  +\frac{r}{\sqrt{\gamma}}
  \frac{\frac{r_0}{\sqrt{\gamma}}}
  {\left(\frac{r_0}{\sqrt{\gamma}}\right)^2+1}
  \right]+\frac{r_0}{\sqrt{\gamma}};
\end{multline}
observe that
\begin{equation}\label{E:throat}
   B\left(\frac{r_0}{\sqrt{\gamma}}\right)
   =\frac{r_0}{\sqrt{\gamma}},
\end{equation}
the analogue of $b(r_0)=r_0$.  Since $B$
is a function of $r$, we may consider the
line element
\begin{equation}\label{E:line2}
ds^{2}=-e^{2\Phi(r)}dt^{2}+\frac{dr^2}
{1-\frac{B(r/\sqrt{\gamma})}{r/\sqrt{\gamma}}}
+r^{2}(d\theta^{2}+\text{sin}^{2}\theta\,
d\phi^{2}).
\end{equation}
It now becomes apparent that in view of
Eq. (\ref{E:throat}), this line element
represents a wormhole with throat radius
$r_0/\sqrt{\gamma}$, while retaining
asymptotic flatness.

To check the flare-out condition, let
\begin{equation}
   x=\frac{r}{\sqrt{\gamma}} \quad \text{and}
   \quad x_0=\frac{r_0}{\sqrt{\gamma}}.
\end{equation}
Then
\begin{equation}
   B(x)=\frac{1}{\sqrt{\gamma}\,}b(x)=\frac{4m}{\pi}
   \frac{1}{r}\left[x\,\text{tan}^{-1}x-
   \frac{x^2}{x^2+1}-x\,\text{tan}^{-1}x_0
   +x\frac{x_0}{x_0^2+1}\right]+x_0.
\end{equation}
So $B(x_0)=x_0$, as before, but we also
have
\begin{equation}
   B'(x)=\frac{1}{\sqrt{\gamma}}\,b'(x)=
   \frac{4m}{\pi}\frac{1}{r}\left[
   \text{tan}^{-1}x+\frac{x}{x^2+1}-
   \frac{2x}{(x^2+1)^2}-\text{tan}^{-1}x_0
   +\frac{x_0}{x_0^2+1}\right]
\end{equation}
and hence
\begin{equation}
   B'(x_0)=\frac{4m}{\pi}\frac{1}{r}
   \left[\frac{2x_0}{x_0^2+1}-
   \frac{2x_0}{(x_0^2+1)^2}\right].
\end{equation}
Since
\begin{equation}
   \frac{x_0}{x_0^2+1}=
   \frac{\frac{r_0}{\sqrt{\gamma}}}
   {\left(\frac{r_0}{\sqrt{\gamma}}\right)^2+1}
   \rightarrow 0\quad \text{as} \quad
   \gamma \rightarrow 0,
\end{equation}
it follows that
\begin{equation}
   B'(x_0)<1 \quad \text{for}\,\, \gamma \,
   \,\text{sufficiently small}
\end{equation}
independently of $m$ and $r$.  (It is
understood that $\gamma$ is always a small
nonzero constant.)  So the flare-out
condition is met.  It also follows that
\begin{equation}
   x_0=\frac{r_0}{\sqrt{\gamma}} \quad
   \text{is macroscopic}.
\end{equation}
It confirms the general discussion in 
Subsection \ref{SS:NG} based on the Einstein
field equations in conjunction with the
noncommutative-geometry background.  By
taking this to be a fundamental property,
the resulting macroscopic scale then becomes
an emergent phenomenon.  This is discussed
further in Section \ref{S:Conclusion}.

\section{A remark on the radial tension}
Given that the radial tension $\tau(r)$
is the negative of the radial pressure
$p_r(r)$, it is noted in Ref. \cite{MT88}
that Eq. (\ref{E:E2}) can be rearranged
to yield
\begin{equation}
   \tau(r)=\frac{b(r)/r-2[r-b(r)]\Phi'(r)}
   {8\pi Gc^{-4}r^2},
\end{equation}
temporarily reintroducing $c$ and $G$.
From this condition it follows that the
radial tension at the throat is
\begin{equation}\label{E:tau}
  \tau(r_0)=\frac{1}{8\pi Gc^{-4}r_0^2}\approx
   5\times 10^{41}\frac{\text{dyn}}{\text{cm}^2}
   \left(\frac{10\,\text{m}}{r_0}\right)^2.
\end{equation}
It is also pointed out in Ref. \cite{MT88}
that for $r_0=3$ km, $\tau$ has the same
magnitude as the pressure at the center
of a massive neutron star. Attributing
this outcome to exotic matter ignores the
fact that exotic matter was introduced to
explain the violation of the NEC.  It is
shown in Ref. \cite{pK20}, however, that
a noncommutative-geometry background can
account for the high radial tension.

\section{Wormhole geometries in $f(Q)$ gravity}
    \label{S:f(Q)}
Our first task in this section estimate of the 
mass of the wormhole.  From Eq. (\ref{E:rho}), 
\begin{equation}
   m(r)=\int^r_{r_0}\rho(r')4\pi (r')^2\,dr'
   =\frac{2m}{\pi}
  \left[\text{tan}^{-1}\frac{r}{\sqrt{\gamma}}
  -\frac{r\sqrt{\gamma}}{r^2+\gamma}
   -\text{tan}^{-1}\frac{r_0}{\sqrt{\gamma}}
  +\frac{r_0\sqrt{\gamma}}{r_0^2+\gamma}
\right].
\end{equation}
Sine $m$ in Eq. (\ref{E:rho}) represents the 
mass of a particle, we conclude that the mass 
$m(r)$ cannot be very large.  It has been shown, 
however, that Morris-Thorne wormholes are 
actually compact stellar objects \cite{pK22}.
The implication is that 
noncommutative-geometry-inspired wormholes are 
likely to be microscopic after all.    

Attempts to overcome the theoretical and practical 
problems confronting Morris-Thorne wormholes have 
relied heavily on various modified gravitational 
theories.  A recently proposed modified theory, 
called $f(Q)$ gravity, is due to Jimenez, et al. 
\cite{JHK18}.  Here $Q$ is the non-metricity 
scalar from the field of differential geometry.  
The action for this gravitational theory is
\begin{equation}
   S=\int \frac{1}{2}f(Q)\sqrt{-g}\,d^4x
   +\int\mathcal{L}_m\sqrt{-g}\,dx^4,
\end{equation}
where $f(Q)$ is an arbitrary function of $Q$, 
$\mathcal{L}_m$ is the Lagrangian density of 
matter, and $g$ is the determinant of the metric 
tensor $g_{\mu\nu}$.  Even though it is a fairly 
new theory, numerous applications have already 
been found; see, for example, Refs. \cite{JHK18, 
zH22, zH21, gM21, uS21, aB21, gM22, fP22, HMSS}.  
Since we are primarily interested in qualitative 
results, our focus is necessarily more narrow.  
Accordingly, we are going to follow Ref. 
\cite{zH22}, in part because it uses the 
commonly employed, but highly idealized, linear 
form $f(Q)=\alpha Q+\beta$.  There is one drawback, 
however: since $f''(Q)=0$ and 
$f'(Q)=\text{a constant}$, this produces the 
Einstein field equations with a cosmological 
constant \cite{lH23}.  Fortunately, $f(Q)$ is 
entirely arbitrary; so consider, for example, 
$f(Q)=\alpha Q^{1+\epsilon}+\beta$, 
$0<\epsilon\ll 1$.  This form is arbitrarily 
close to $f(Q)=\alpha Q+\beta$, while avoiding the 
above drawback.  In other words, we can view 
$f(Q)=\alpha Q+\beta$ as a convenient approximation 
that is sufficient for present purposes since it 
readily provides us with the desired qualitative 
result, as we will see.  Returning to Ref.
\cite{zH22}, the corresponding field equations are:
\begin{equation}\label{E:Einstein1}
   \rho=\frac{\alpha b'}{r^2}+\frac{\beta}{2},
\end{equation}
\begin{equation}
   p_r=\frac{1}{r^3}\left[2\alpha r(r-b)
   \Phi'-\alpha b\right]-\frac{\beta}{2},
\end{equation}
and
\begin{equation}
   p_t=\frac{1}{2r^3}\left[\alpha(r\Phi'+1)
   (-rb'+2r(r-b)\Phi'+b\right]+
   \frac{\alpha(r-b)\Phi''}{r}-\frac{\beta}{2}.
\end{equation}
Eq. (\ref{E:Einstein1}) can now be combined with 
Eq. (\ref{E:rho}):
\begin{equation}
   \frac{m\sqrt\gamma}{\pi^2(r^2+\gamma)^2}=
   \frac{\alpha b'(r)}{r^2}+\frac{\beta}{2}.
\end{equation}
Solving for $b'(r)$, we get
\begin{equation}
   b'(r)=\frac{1}{\alpha}
   \frac{mr^2\sqrt{\gamma}}{\pi^2(r^2+\gamma)^2}
   -\frac{1}{2}\frac{\beta}{\alpha}r^2
\end{equation}
and
\begin{equation}
   b(r)=\frac{1}{\alpha}\int^r_{r_0}
   \left[\frac{m(r')^2\sqrt{\gamma}}{\pi^2[(r')^2+\gamma]^2}
   -\frac{1}{2}\frac{\beta}{\alpha}(r')^2 \right]\,dr'+r_0.
\end{equation}
To retain asymptotic flatness, we let $\beta =0$, yielding
\begin{equation}\label{E:shapefunction}
   b(r)=\frac{1}{\alpha}\frac{m\sqrt{\gamma}}{\pi^2}
   \left[\frac{\text{tan}^{-1}\frac{r}{\sqrt{\gamma}} }{2\sqrt{\gamma}}
   - \frac{r}{2(r^2+\gamma)}-                 
   \frac{\text{tan}^{-1}\frac{r_0}{\sqrt{\gamma}} }{2\sqrt{\gamma}}
   +\frac{r_0}{2(r_0^2+\gamma)}
    \right] +r_0. 
\end{equation}
Observe that $b(r_0)=r_0$, as required.  Thanks 
to the free parameter $\alpha$ from $f(Q)$ gravity, 
the mass of the wormhole, 
$m(r)=\int^r_{r_0}\rho(r')4\pi (r')^2\,dr'
=\frac{1}{2}b(r)$ from Eq. (\ref{E:E1}), 
can now be macroscopic.  This is our main 
conclusion.

\emph{Remark:} Similar comments can be made 
about wormholes whose energy violation is due 
to the Casimir effect, also discussed in Ref. 
\cite{zH22}.  Here the energy density is 
$-\hbar c\pi^2/720r^4$, so that
\begin{equation}
   \rho(r)=\frac{\alpha b'(r)}{r^2}+
   \frac{\beta}{2}
   =-\frac{\hbar c\pi^2}{720r^4}.   
\end{equation}
Letting $\beta=0$ again, solving for 
$b'(r)$, and then integrating, we obtain 
\begin{equation}
   b(r)=\frac{1}{\alpha}\frac
   {\hbar c\pi^2}{720}\left(\frac{1}{r}
   -\frac{1}{r_0}\right)+r_0,
\end{equation}
showing that $m(r)=\frac{1}{2}b(r)$ can 
be macroscopic. 

\section{Discussion and Conclusions}
   \label{S:Conclusion}
The goal of this paper is to obtain a viable
model for a macroscopic traversable wormhole
by starting with a noncommutative-geometry
background involving a particular microscopic
effect, the replacement of  point-like
particles by smeared objects.  In the usual
terminology, this is an example of a
\emph{fundamental property}.  Applied to
wormholes, one can make use of the following
assertions discussed in Ref. \cite{NSS06}:
noncommutative effects can be implemented in
the Einstein field equations by modifying
only the energy momentum tensor, while
leaving the Einstein tensor unchanged, thereby
implying that the length scales could be
macroscopic.

Returning to the smearing effect,  this is
expressed in Eq. (\ref{E:rho}) in mathematical
form, now seen as a fundamental property.
Eq. (\ref{E:rho}) has led to the shape function
$B(x)$, which, in turn, results in a macroscopic \
throat radius.  This outcome is a typical
example of an \emph{emergent} property since
it does not appear in the fundamental theory.
(The concept of emergence dates at least from
the time of Aristotle.)  By definition, emergent
properties or objects are derived from a
fundamental theory.  Such a process is not
reversible, however: in our case, the
macroscopic scale does not yield the
smearing effect in the original theory.  The
result is an \emph{effective model} for
a macroscopic wormhole precisely because the
short-distance effect has been discarded:
not only is this information no longer
needed, it is meaningful only in the
fundamental theory. 

It is shown in this paper that such 
wormholes cannot have a large enough 
mass to exist on a macroscopic scale.  
However, Morris-Thorne wormholes are 
likely to be compact stellar objects, 
akin to neutron stars, and so would 
normally be quite massive.  By invoking 
$f(Q)$ modified gravity, it is shown
that the resulting extra degrees of 
freedom enable us to overcome these 
obstacles, thereby allowing the wormholes  
to be sufficiently massive despite the 
noncommutative-geometry background.
This conclusion also pertains to 
Casimir wormholes. 
\\
\\
\textbf{Acknowledgment:} The author
  would like to thank Dr. Josiah Yoder
  for the helpful discussions on the
  topic of emergence.
\\
\\
\noindent
CONFLICTS OF INTEREST
\\
\noindent
The author declares that there are no
conflicts of interest regarding the
publication of this paper.

\end{document}